# Magnetic ordering of the RE lattice in REFeAsO: the odd case of Sm. A specific heat investigation in high magnetic field.

S. Riggs, C. Tarantini, J. Jaroszynski, A. Gurevich

National High Magnetic Field Laboratory 1800 East Paul Dirac Drive, Tallahassee, FL 32310 USA

A. Palenzona, M. Putti

CNR-INFM-LAMIA and Università di Genova via Dodecaneso 33, 16146 Genova Italy

T. Duc Nguyen and M. Affronte.

CNR-INFM-S3 National Research Centre and University of Modena and Reggio Emilia, via G. Campi 213A, 41100, Modena, Italy

**Abstract.**

We have investigated the evolution of the low temperature specific heat anomaly ($T_N$=5.4K in zero field) in polycrystalline SmFeAsO samples with magnetic fields up to 35T. The anomaly remains very sharp up to 16T and becomes rounded with little shift in temperature at higher fields. Doped (superconducting) SmFeAsO$_{0.85}$F$_{0.15}$ sample shows a similar behavior up to 16T. The initial slope of the critical field $dB_c/dT$ is 160T/K for undoped SmFeAsO and 70T/K for doped SmFeAsO$_{0.85}$F$_{0.15}$, with $B_c(T)$ defined at the peak of the specific heat anomaly. The insensitivity to the application of an external magnetic field is unique to Sm and is not observed in CeFeAsO whose anomaly shifts with initial slope $dB_c/dT$=5.7T/K. We argue that SmFeAsO(F) presents an unprecedented case of spin reorientation at the antiferromagnetic transition.

PACS numbers: 75.40.Cx, 75.30.Kz, 75.50.Ee, 74.70.Dd

**Introduction.**

The recently discovered oxypnictides are extremely rich materials in which different phenomena, namely spin density waves (SDW), superconductivity and magnetic order, have been observed. So far, most of the attention has been devoted to the interplay between spin density waves (SDW) and superconductivity in this family of compounds. Antiferromagnetic (AFM) order of the rare earths (RE) sublattice has been observed at liquid helium temperature in REFeAsO derivatives with magnetic RE= $Sm^1$, $Nd,^2$ $Ce,^3$ and $Pr.^4$ This static magnetic order is little affected by charge doping of the FeAs conducting layers and it coexists with superconductivity. Magnetic order of the RE sublattice and superconductivity have been observed in heavy fermion systems where it has been established that the interplay between RE and free electrons is the origin of peculiar phenomena in these materials. Current research is aimed at understanding the nature of similar phenomena in the oxypnicitides. AFM order of the RE sublattice have been observed in also the $RE_2CuO_4$ cuprates which share similar structure and properties with the oxypnictides. Both systems are layered structures where the conducting layers become superconducting upon doping. Separating the conducting layers are the RE sheets where the RE ions order antiferromagnetically. The layered structure induces an anisotropic behavior and introduces a hierarchy of magnetic interactions that may lead to an intriguing magnetic phase diagram. In cuprates, RE are very sensitive probes of the crystalline environment and show subtle differences between one RE to another.$^5$ For instance, the RE ground state splits in different ways depending of the point symmetry of the crystal and it is quite sensitive to the presence of localized Cu spins$^5$. Interaction between superconductivity and AFM has been reported for doped $Sm_2CuO_4$.$^{6,7}$ In the oxypnictides, an interplay between Fe and RE ordered moments have been reported for RE=$Nd^2$, $Ce^8$ and $Pr^4$ while neutron diffraction experiments on SmFeAsO are still lacking. Direct comparison of different behaviors shown by REFeAsO with different RE can thus reveal important details on the physics of these materials.

This work started from some preliminary investigations we performed on SmFeAsO samples that showed that the specific heat anomaly related to the ordering of the Sm sublattice is almost insensitive to the application of an external magnetic field up to 7T. This seems inconsistent with the relatively low critical temperature $T_N$=5.4K and the conventional molecular field theory. It was immediately evident that the case of Sm is odd within the family of REFeAsO since, for instance, the magnetic transition in CeFeAsO is more sensitive to the application of moderate fields.$^9$ Interestingly, $Sm_2CuO_4$ cuprate exhibits AFM ordering of $Sm^{3+}$ ions at $T_N$=5.94 K and a similar insensitivity to the application of high magnetic field was claimed$^{10}$.

The aim of this work is primarily to provide new experimental information on the magnetic ordering of Sm sublattice. We performed specific heat measurements up to 35T and monitor the evolution of the AFM transition in polycrystalline SmFeAsO. To our knowledge, very few –if any- experiments have been performed so far onto an antiferromagnet at such high fields. Our experiments reveal a surprising insensitivity of the antiferromagnetic transition to the application of an external magnetic field. These results are compared to those obtained on doped SmFeAsO$_{0.85}$F$_{0.15}$ and on CeFeAsO. We argue that the SmFeAsO presents an unprecedented case of spin reorientation transition and this may also have consequences for superconductivity.

**Experimental results.**

Polycrystalline samples were prepared in two steps as described in ref.[11] : 1) synthesis of REAs from pure elements; 2) synthesis by reacting REAs with stoichiometric amounts of Fe, Fe$_2$O$_3$ and FeF$_2$ at high temperature. Samples were characterized by X-ray powder diffraction followed by Rietveld refinement, revealing their single-phase, high crystalline nature. TEM analysis evidences the lack of structural defects. Starting from a larger bulk sample, pieces of few tenth of mgr were cut in a parallelepiped shape for specific heat measurements.

Heat capacity up to 16T was measured by Quantum Design PPMS using the two-tau method. Measurements up to 35T were performed at NHMFL in Tallahasse by using the relaxation technique on an in-house custom built calorimeter calibrated for high magnetic fields.

Elsewhere[12] we reported the analysis of specific heat data from ~10K to 300K by considering an electronic $C_{el}=\gamma T$ and a lattice contribution $C_{latt}$ which, in turn, includes both Debye and Einstein terms. The addition of magnetic Schottky anomalies, as suggested by Baker et al. [13], may further refine this data analysis. In Table 1 we summarize the main parameters we extracted from the analysis of data obtained on our samples.

Table 1. Essential electronic, magnetic and lattice parameters extracted from the analysis of specific heat data.

| Sample | $T_N$ (K) | $\gamma$ (mJ/molK$^2$) | Debye temperature $\Theta_D$(K) | Einstein temperatures $\Theta_E$(K) | Schottky $\Delta/k_B$ (K), |
|---|---|---|---|---|---|
| SmFeAsO | 5.4 | 42±2 | 190±5 | 201, 405 | 266, 654 |
| SmFeAsO$_{0.85}$ F$_{0.15}$ | 3.75 | 44±2 | 170±5 | 141, 254, 409 | - |
| CeFeAsO | 3.9 | 36±2 | 188±5 | 220, 550 | 216, 785 |

We focus here on anomalies at low temperature. The specific heat *C(T,B)* of undoped SmFeAsO is plotted in Fig.1 as a function of temperature (T) for different applied magnetic fields (B). The anomaly looks extremely sharp in zero field, with jump ΔC as high as 20J/molK. No thermal hysteresis can be observed by consecutive cooling and warming of the sample. Data analysis (reported in ref.12) shows that the magnetic entropy tends to saturate to Rln2 consistent with a dublet ground state of the $Sm^{3+}$ ions. This also indicates that the bulk of the sample is involved in this ordering process, thus confirming the good quality of our sample.

In fig.1 two sets of data, independently taken by a PPMS (up to 16T) and with the home-made calorimeter at NHMFL (from 20 to 35T) on a SmFeAsO polycrystalline sample, are plotted in the same graph: the two data sets scale smoothly one on top of the other demonstrating excellent reproducibility of results, also considering that different setups have been used in different experiments. The most striking feature of these data is that the specific heat anomaly is perturbed very little by the application of magnetic field: at 16T the peak is shifted by only 0.2K and the *C(T, B=16T)* anomaly is still very sharp. Stronger magnetic fields progressively make the peak more rounded.

For comparison, the same anomaly was measured on a doped superconducting polycrystalline $SmFeAsO_{0.85}F_{0.15}$ sample (fig.2) and an undoped CeFeAsO polycrystalline sample (fig.3). For doped $SmFeAsO_{0.85}F_{0.15}$, $\Delta C \approx 5 J/molK$, and the zero-field peak at 3.70K shifts down to 3.45K in a magnetic field of 16T. Surprisingly, the anomaly in both doped and undoped SmFeAsO does not get much broader under the application of magnetic fields up to 16T. Conversely, the anomaly in undoped CeFeAsO sample (fig.3) is more sensitive to an external magnetic field: the peak shifts from 3.9K to 3.1K with only 5T and the anomaly clearly gets broader in 7T. After subtraction of electronic and lattice contribution, estimation of magnetic entropy *S* for CeFeAsO shows that *S* tends to saturate to 0.5R above 5K, definitively a lower value as compared to Rln2 for SmFeAsO. This is essentially due to the different split of the ground multiplets in the two compounds as discussed below. The broadening of the specific heat anomaly in CeFeAsO suggests that the magnetic entropy tends to saturate at the same value for dfferent magnetic fields while this does not seems to be the case for the Sm- samples.

To get more insight on this phase transition, specific heat was measured at zero field with very small heat pulses in order to approach the transition more closely. After subtraction of the background contribution $C_0 = \gamma T + \beta T^3$ with ($\gamma = (42 \pm 2) mJ/molK^2$) and ($\beta = (0.36 \pm 0.04) mJ/molK^4$) of ref. 12, the analysis of the fluctuation part $\delta C(T) = C(T) - (\gamma T + \beta T^3)$ was performed. Fluctuations above (+) and below (-) $T_N$ are expected to scale with the reduced temperature $t = |T_N - T|/T_N$ as $\delta C^{(+/-)} = A^{(+/-)} t^\alpha$ where $T_N$, $A^{(+/-)}$ are materials parameters while the critical exponents $\alpha^{(+/-)}$, determined by the universality class of the phase transition, must be the same above and below $T_N$. By plotting $\log(\delta C^{(+/-)})$ as a function of $\log(t)$ separately for data above and below $T_N$, we obtained the graph shown in fig.4. It is generally accepted that the true critical exponents are obtained by getting extremely close to $T_N$ and this requires excellent sample homogeneity and extremely small heat pulses during the heat capacity measurements. We performed a special run of measurements in zero field with $t<0.2\%$ which is essentially limited by the accuracy on the reduced temperature $t$ we can experimentally achieve. Two parallel straight lines, one of which extends over almost two decades, can be obtained by choosing $T_N$ =5.392K, which, in turn, is very close to the maximum of the *C(T)* peak. From this analysis we obtained the critical exponents $\alpha^+=\alpha^-=0.316\pm0.01$. Peaks in doped $SmFeAsO_{0.85}F_{0.15}$ and undoped CeFeAsO are more rounded therefore the same analysis can not be extended close enough to $T_N$ to get reliable values of the critical exponents. In the case of $SmFeAsO_{0.85}F_{0.15}$ this problem may be due to the sample inhomogeneities making the AFM transition less sharp. Conversely, CeFeAsO sample is homogeneous and even more crystalline than SmFeAsO sample, as revealed by structural and microstructural analyses, and the different shape of the peaks is probably a further signature of the diverse nature of the AFM transition in the two compounds.

**Discussion.**

Ordering of the RE sublattice have been observed in different oxypnictides, namely REFeAsO derivatives with RE=Ce, Nd, Pr, Sm. Interestingly, there are many similarities between oxypnictides and cuprates. In particular, AFM transitions at liquid helium temperatures have been observed in both families of compounds. $RE_2CuO_4$ have been largely studied in the last fifteen years[5] so it is worth starting with a review of the results obtained in these mirror compounds.

$Sm^{3+}$ ($Ce^{3+}$) have five (one) electrons, respectively, in their 4*f* shell. According to Hund's rules, the resulting free ion electronic ground state is $^6H_{5/2}$ ($^2F_{5/2}$), i.e. a sixfold degenerate level with total angular momentum of J=5/2. In $Sm_2CuO_4$, $Sm^{3+}$ ions are located on sites with $C_{4v}$ symmetry. At high temperature SmFeAsO has tetragonal crystallographic structure with $Sm^{3+}$ in $C_{4v}$ point symmetry, similarly to the case of $Sm_2CuO_4$. A low temperature SmFeAsO phase has an orthorombic structure with $Sm^{3+}$ in a $C_{2v}$ environment. In this case $Sm^{3+}$ has four identical bonds with oxygen and four more bonds with As, two of which differ by only 0.01Å from the other two[14].

Group theory predicts that a crystalline electric field with tetragonal symmetry splits the sixfold degenerated ground-state multiplet into three doublets[15]. Crystalline electric field (CEF) effects have been calculated by Sachidanandam et al.[16] and by Strach et al.[17] for $Sm^{3+}$ in the $C_{4v}$ environment of $Sm_2CuO_4$. The J=5/2 multiplet is split in three dublets, with *a*|5/2>-*b*|-3/2> ground state, a first |1/2> excited state separated by 18.15meV and a ( *b*|5/2>+*a*|-3/2>) dublet 37.76meV at higher energy. Analysis of specific heat by Baker et al.[13] has shown that the pattern of the J=5/2 multiplets is similar for SmFeAsO with a first excited dublet at 22.92meV and a further dublet at 56.4meV from the ground state.

Similarly, CEF effects have been estimated for $Ce^{3+}$ in CeFeAsOF by Chi et al.[8]. In undoped CeFeAsO, $Ce^{3+}$ has local point symmetry $C_{2v}$ and the CEF levels have three magnetic doublets in the paramagnetic state, with $|1/2>$ ground state and $(-c|-5/2>+d|3/2>)$ and $(d|-5/2>+c|3/2>)$ excited states at 18.7eV and 67.7meV respectively. These doublets split into six singlets when the Fe ions order (around 150K). Although deeper investigations are necessary to elucidate the actual split of $Sm^{3+}$ levels oxypnictides, the above mentioned results depict the ground multiplet of $Sm^{3+}$ and $Ce^{3+}$ ions in the relative compounds.

Another peculiarity that distinguishes $Sm^{3+}$ from all other rare earths is its special uniaxial magnetic anisotropy with an easy axis along the crystallographic *c*-axis as was discussed for $Sm_2CuO_4$[16]. Experimental evidences for this in SmFeAsO are still missed due to the lack of large single crystals but, due to the similarities between the two compounds, a similar uniaxial anisotropy can be assumed for SmFeAsO as well and the respective term $H_{an}$ should be considered for the spin Hamiltonian describing this magnetic system.

The ordered magnetic moment of $Ce^{3+}$ in CeFeAsO has been evaluated by neutron diffraction giving $0.83\mu_B$[2]. This estimate is not available for $Sm^{3+}$ in SmFeAsO, but carrying on the analogy with $Sm^{3+}$ in $Sm_2CuO_4$, we can assume it to be about $0.37\mu_B$ as estimated by neutron diffraction for $Sm_2CuO_4$ [19]. Note that within the simplest molecular field approach, a larger magnetic moment for $Ce^{3+}$ should imply a higher $T_N$ in comparison with $Sm^{3+}$. Experimentally this is not the case, in fact $T_N$ for SmFeAsO is higher than for CeFeAsO.

Magnetic coupling between $Sm^{3+}$ ions may result from two contributions: short range in-plane interaction $H_{inplane}$ of probably a predominantly ferromagnetic super-exchange origin, and long range RKKY type interplane AFM interaction $H_{LR}$, mediated by conducting electrons on the FeAs planes. It was noticed that SmFeAsO compounds have relatively high $\gamma$ values[1,12] and Pauli susceptibility[18], which suggests some hybridization/interaction of conducting electron with the $Sm^{3+}$ magnetic ion and collective (Stoner) renormalization effects. Moreover, it should be noticed that charge doping in FeAs planes affects, but does not drastically change the magnetic ordering of the Sm lattice, which again is quite similar to what happens in $Sm_2CuO_4$. However, CeFeAsO shows $\gamma$ value close to those for SmFeAsO (see table 1) and behaves in a similar way after doping, thus the interaction of the RE with conducting electrons cannot be considered as the distinctive feature between the two compounds.

In $Sm_2CuO_4$, inelastic neutron diffraction experiments[19] have shown that the Sm lattice undergoes an AFM transition to the structure comprised of ferromagnetic sheets within the *ab* planes in which the Sm spins align along the c-axis and alternate their direction between neighboring layers, a unique case within the $RE_2CuO_4$ family.[5] For SmFeAsO we expect a similar type of order, which is also unique in the family of REFeAsO pnictides where $Ce^{3+}$ and $Nd^{3+}$ order antiferromagnetically with the spins along the *ab*-planes in CeFeAsO[2] and NdFeAsO[3] respectively.

Within this scenario, we may interpret our results. In spite of the relatively low critical temperature, a very strong magnetic field is required to break this ordering. Up to 16T, the anomalies in *C(T)* are quite sharp and we may evaluate the initial slope of the critical field $dB_c/dT$ just considering the shift of the specific heat peak. This yields $dB_c/dT$=160T/K and 70T/K for undoped SmFeAsO and doped SmFeAsOF, respectively. Such steep slopes are quite surprising given the Nèel temperature of ≈5K, and if compared with $dB_c/dT$=5.7T/K observed on CeFeAsO. High critical fields are typical for antiferromagnets with spin flip (flop) transitions and similar cases have been reported for heavy fermion systems [20, 21, 22, 23, 24, 25], quasi 2D antiferromagnet with triangular lattice[26] and pyroclore structures[27]. The case of SmFeAsO remains, however, unprecedented for the very small effects caused by rather strong magnetic fields. We believe that such insensitivity of the transition to strong magnetic fields could be due to the small magnetic moment of $Sm^{3+}$, and, more importantly, to the uniaxial anisotropy.

In spite of the experimental difficulties (lack of single crystals, need of strong magnetic fields) the behavior of fig.1 may reveal an intriguing case for a spin reorientation (spin flip/flop or metamagnetic[15]) transition in an antiferromagnet. The external magnetic field competes with two types of coupling ($H_{inplane}$ within the Sm planes and $H_{LR}$ between Sm planes) and with the single ion magnetic anisotropy ($H_{an}$). This situation commonly leads to multicritical points with a variety of metastable magnetic configurations due to the interplay between different energies[28].

The modality in which the external field reorients spins depends on the field orientation with respect to the direction of the magnetic order (represented by the Nèel vector in an antiferromagnet). For fields parallel to the Nèel vector (c-axis in our model), there is an abrupt reversal of the magnetization in alternated planes, i.e. a spin flip transition. For external magnetic field perpendicular to the Nèel vector, there is a progressive canting of the magnetic moments along the direction of the magnetic field. In our case we used polycrystalline samples so the external magnetic field probes all possible directions of magnetization of Sm sublattices. At the highest fields (≥20T) the specific heat anomaly gets broader. The broadening partially reflects the polycrystalline nature of the sample (in which local transitions occur in the grains with the *ab*-planes parallel to **B**) but it may also be due to the appearance of a metastable magnetic phase. In this case first order (metamagnetic) phase transition and different entropy balance are expected[28]. Within our experimental accuracy, we did not detect any latent heat but more accurate experiments are required to clarify this issue.

In fig.1, we may identify a low temperature edge that progressively shifts towards lower temperature while the almost unperturbed kink of the *C(T)* curve at 5.5K indicates that full saturation of the magnetization in all directions requires huge external fields in SmFeAsO. This suggests that the uniaxial anisotropy can be very high. If we define thecritical field $B_c$ corresponding to the low temperature knee of the transition we find that $B_c(T)$ approximatively follows a power law $B_c(T) \propto (T_N - T)^\beta$ with an exponent β≈0.5 for $T>0.2T_N$, as shown in Fig. 5. Within a simple mean field approach, this behaviour is expected for spin flip transition[29] and it has been reported for a $Sm_2CuO_4$ single crystal[30]. Single crystal experiments will better clarify the actual magnetic phase diagram.

Analysis of thermal fluctuations provides a critical exponent α=0.316±0.01 in zero field. According to the previous discussion, we expect the magnetic order in the Sm lattice to be a 3D Ising system (the ground state of $Sm^{3+}$ being a dublet with uniaxial anisotropy). In the case of purely short range interaction the expected critical exponent α ranges between 0.1 and 0.15 [28]. The discrepancy we found may be due to long range interaction ($H_{LR}$) between the Sm planes.

The behavior of doped $SmFeAsO_{0.85}F_{0.15}$ is similar to what observed in undoped SmFeAsO and it can therefore be interpreted in the same framework. The main difference is the increase of the carrier density in the FeAs planes that probably changes the magnetic coupling ($H_{LR}$) between the Sm planes. The relevant point is, however, that the AFM order of the Sm sublattice is not dramatically affected by the disappearance of the spin density waves in the FeAs planes and it coexists with superconductivity. This leads to an interesting question: what is the interplay between the AFM order in the Sm sublattice and superconductivity?

Evidence of interplay between superconductivity and AFM have been reported for electron doped $(SmCe)_2CuO_4$: the temperature dependence of the upper critical field, $H_{c2}$, on a sample with critical temperature $T_c$ = 11.4 K displays an anomalous upturn at $T/T_c$ =0.5, just in the vicinity of the Sm ordering temperature [6]. Moreover, penetration depth measurements indicate a spin-freezing transition that dramatically increases the superfluid density below $T_N$ [7]. Recently Lake et al.[31] have shown that in layered superconductors $(LaBa)_2CuO_4$ AFM order actually coexists with superconductivity and it may directly affect the mixed state by straightening vortex lines. Huge $H_{c2}$ values of $SmFeAs(OF)$[32] do not allow to investigate it at low temperature. However, torque measurements[33] in the mixed state of $SmFeAsO_{0.8}F_{0.2}$ single crystal have shown an anomalous increase of the anisotropy factor starting at 20K. This could be related to the incipient AFM transition in Sm lattice and interplay between vortex lines and incipient AFM order is taking place. The combination of huge $H_{c2}$ and insensitivity of the AFM transition might have important consequences for applications that deserve further attention.

In summary, we have performed specific heat measurements on polycrystalline SmFeAsO sample up to 35T in order to investigate the magnetic transition involving the Sm sublattice. The observed evolution of the specific heat anomaly in SmFeAsO reveals a surprising insensitivity of the Nèel temperature to the application of strong magnetic fields that survives upon charge doping in SmFeAsO$_{0.85}$F$_{0.15}$ but it is not present in CeFeAsO. Comparing our results to the mirror Sm$_2$CuO$_4$ compounds we argue that the peculiarity of the Sm-based oxypnictides observed in this work is related in part to the small magnetic moment of Sm$^{3+}$ and mostly to the uniaxial magnetic anisotropy.

**Acknowledgements.**
A portion of this work was performed at the National High Magnetic Field Laboratory, which is supported by NSF Cooperative Agreement No. DMR-0654118 by the State of Florida, and by the DOE. This work was also partially supported by the Italian Foreign Affairs Ministry (MAE) - General Direction for the Cultural Promotion and by CNR under the project Short Term Mobility. We are pleased to thank F. Canepa, R. Cimberle, A. Martinelli, M. Tropeano (University of Genoa) for stimulating discussion.

**References.**

**Figure Caption.**

**Fig.1** Specific heat *C(T,H)* anomaly in undoped SmFeAsOF for different magnetic field strengths. A critical field $B_c$ can be defined at the maximum of peak for fields B<16T. For higher fields we arbitrarily extend the $B_c$ definition to the low temperature edge of the *C(T,H)* curves, as indicated by the arrow for the *C(35T)* curve. The start of the anomaly is also evident at high temperature (around 5.5K) and this allows to identify the beginning of the transition.

**Fig.2** Specific heat *C(T,H)* anomaly in polycrystalline SmFeAsO$_{0.85}$F$_{0.15}$ sample in magnetic field up to 16T.

**Fig. 3** Specific heat anomaly in polycrystalline CeFeAsO sample.

**Fig. 4** Thermal fluctuations observed in the specific heat anomaly above and below $T_N$ in an undoped SmFeAsO polycrystalline sample. Data analysis, explained in the text, allows to determine $T_N$=5.392K.

**Fig. 5** Tentative magnetic phase diagram relative to the RE sublattice in SmFeAsO. The grey area indicates where the (Sm) spin reorientation occurs upon the application of an external magnetic field oriented in all possible crystallographic directions. The critical field $B_c$ (squares) is defined as the low temperature knee in the specific heat *C(T,H)* anomaly while the beginning of the transition is indicated by circles (see also fig.1). Note the steep initial slope of the critical field $dB_c/dT$ that we evaluated as high as 160T/K for SmFeAsO.

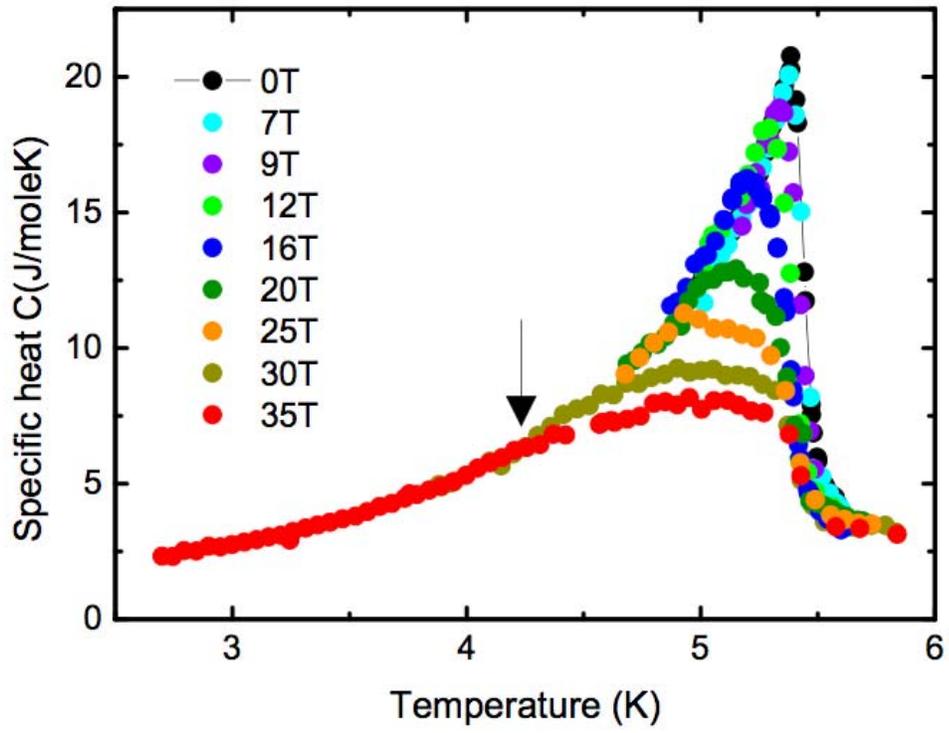

Figure 1

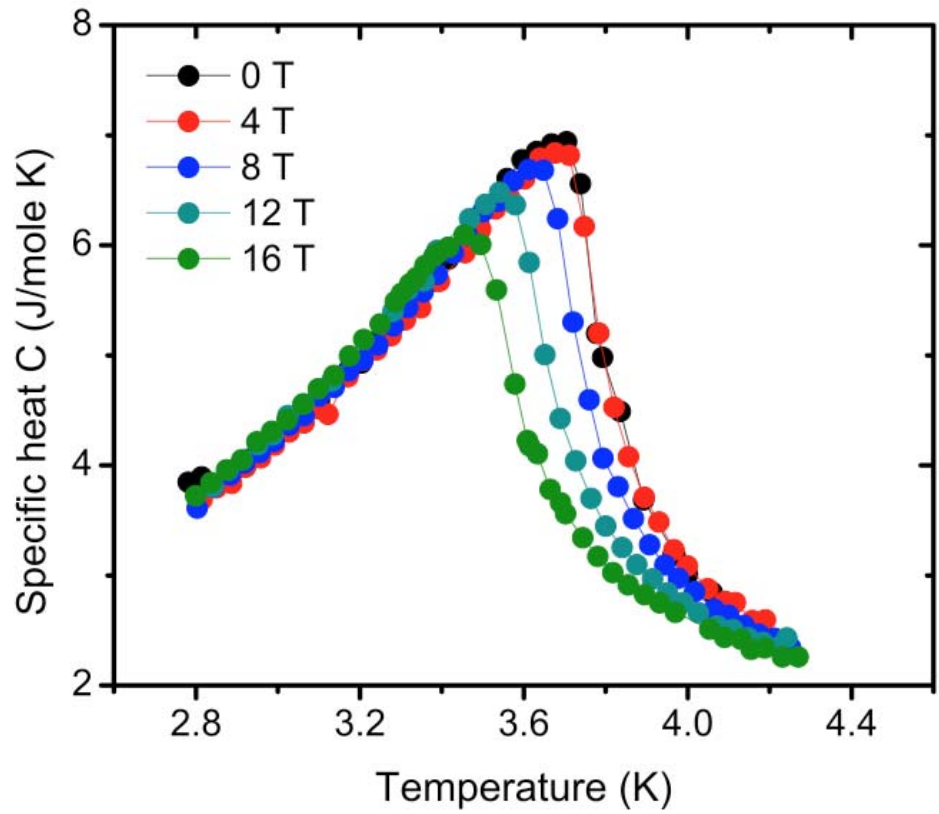

Figure 2

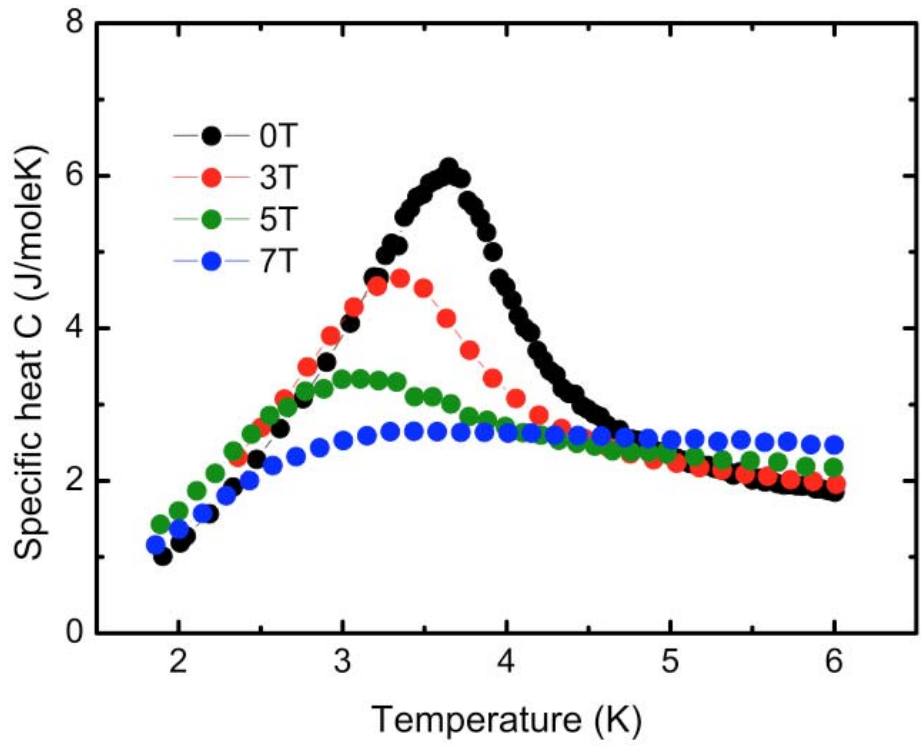

Figure 3

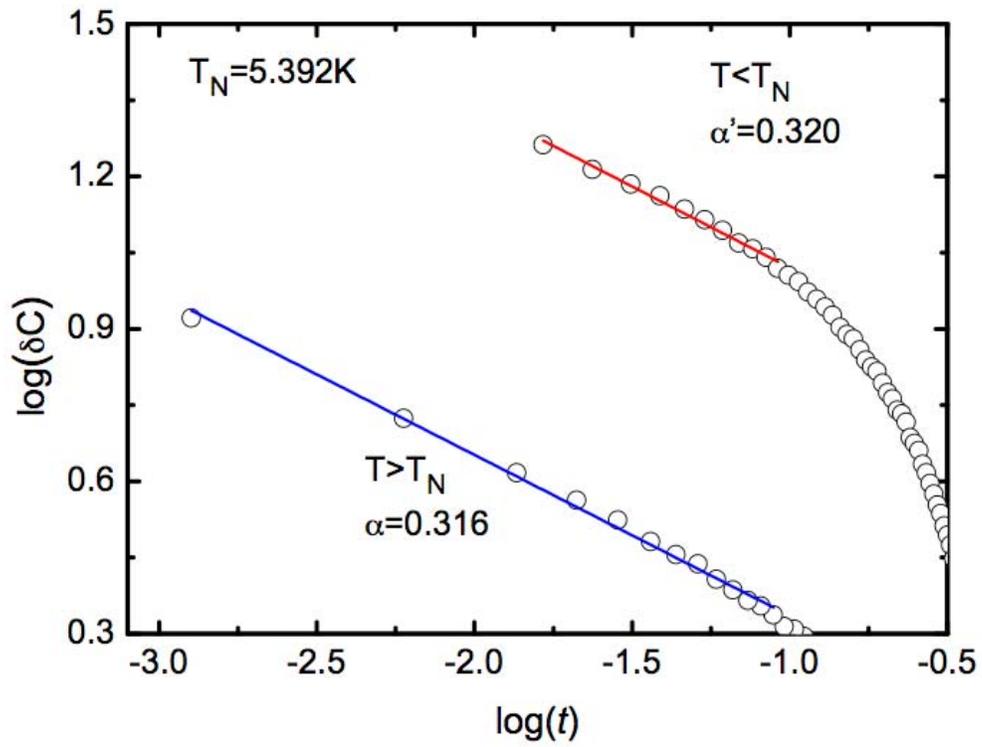

Figure 4

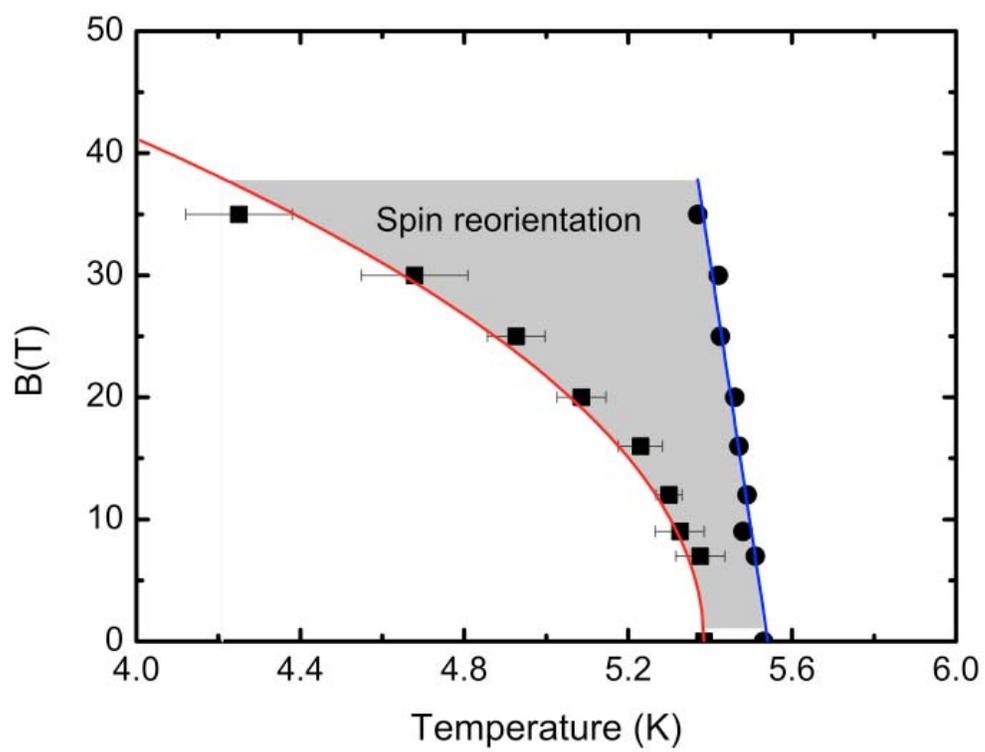

Figure 5